\documentclass[12pt,a4paper,english]{elsarticle}
\usepackage[T1]{fontenc}
\usepackage[latin9]{inputenc}
\usepackage{units}
\usepackage{amstext}
\usepackage{amssymb}
\usepackage{setspace}
\usepackage{esint}
\makeatletter

\pdfpageheight\paperheight
\pdfpagewidth\paperwidth

\@ifundefined{date}{}{\date{}}
\journal{International Journal of Modern Physics D}

\makeatother

\usepackage{babel}
\begin{document}

\title{''Microscopic'' approach to the Ricci dark energy}

\author{Bogus\l{}aw Broda}

\ead{bobroda@uni.lodz.pl}

\ead{http://merlin.fic.uni.lodz.pl/kft/people/BBroda}

\address{Department of Theoretical Physics, University of \L{}\'od\'{z}, Pomorska
149/153, PL--90-236~\L{}\'od\'{z}, Poland}
\begin{abstract}
A derivation of the Ricci dark energy from quantum field theory of
fluctuating {}``matter'' fields in a classical gravitational background
is presented. The coupling to the dark energy, the parameter $\alpha$,
is estimated in the framework of our formalism, and qualitatively
it appears to be within observational expectations.\end{abstract}
\maketitle
\def\citet{\cite}

\noindent \begin{flushleft}
The Ricci dark energy (DE) model (proposed in \citet{Gao:2007ep})
is a model of DE with the energy density proportional to the Ricci
scalar with the {}``coupling constant'' $\alpha$ \citet{All}.
The Ricci DE model is inspired by, or is a sort of, holographic models.
The Ricci DE density is defined by
\begin{equation}
\varrho_{\textrm{d}}=\frac{3\alpha}{\kappa}\left(\dot{H}+2H^{2}\right),\label{eq:defroD}
\end{equation}
where $H$ is the Hubble parameter ($H\equiv\dot{a}/a$), $\kappa\equiv8\pi G$
with the Newton gravitational constant $G$, and $\alpha$ is an \emph{a
priori} free coupling constant. The general form of the DE density
(\ref{eq:defroD}) is postulated rather \emph{ad hoc} (or at best
using qualitative arguments), and is in a sense of a {}``phenomenological''
origin. The term {}``phenomenological'' means that nor a definitive
value of the constant $\alpha$, nor even the form (\ref{eq:defroD})
is derived from a fundamental theory, e.g.\ from quantum field theory.
Therefore the constant $\alpha$ is a free parameter, as usually in
the framework of phenomenological approaches (for experimental estimations
of $\alpha$ see, e.g., \citet{Wang:2011km}). In \citet{Broda} we
have proposed a model of DE that is derived from vacuum quantum fluctuations
of {}``matter'' fields in the framework of the formalism of quantum
field theory. But it was before the advent of the Ricci DE models.
In the meantime we have realized that actually our model excellently
fits the general scheme of the Ricci DE models. Since our derivation
is definitely {}``microscopic'' rather than {}``phenomenological'',
we are able to estimate the constant $\alpha$ as well. In fact, it
is possible not only to estimate the constant $\alpha$, but also
to show that qualitatively it is within expected range from observational
point of view. Using arguments coming from quantum field theory of
{}``matter'' fields in a classical gravitational background, we
have calculated the contribution to the density of DE coming from
a single particle mode (for the Euclidean version see \citet{Broda}).
Namely, an effective quantum contribution coming from a single {}``matter''
mode is represented by the functional determinant $\det i\mathcal{D}$,
where $\mathcal{D}$ is a second-order differential operator in a
classical background of the gravitational field $g_{\mu\nu}$. In
terms of determinants we have the effective action
\begin{equation}
S_{\textrm{eff}}=\pm\frac{i}{2}\log\det i\mathcal{D},\label{eq:logdetd}
\end{equation}
where the sign {}``$+$'' or {}``$-$'' corresponds to bosons
or fermions, respectively. For our purposes, the best suited expansion
of (\ref{eq:logdetd}) is the heat-kernel expansion of the Schwinger
proper-time representation of (\ref{eq:logdetd}) \citet{DeWitt}.
Since the ''cosmological constant'' (or DE) should correspond, as
is commonly expected, to the first, {}``constant'' term of the expansion,
i.e.\ to $\int\sqrt{-g}\, d^{4}x$, we will limit our considerations
only to that term. Besides, that terms is dominant (the highest power
of the UV momentum cutoff $\Lambda$). Therefore
\begin{equation}
S_{\textrm{eff}}^{\Lambda}=\mp\frac{i}{2}\intop_{\Lambda^{-2}}^{\infty}\frac{ds}{s}\textrm{Tr}e^{-is\mathcal{D}},\label{eq:SeffHeat}
\end{equation}
where
\begin{equation}
\textrm{Tr}e^{-is\mathcal{D}}=-\frac{i}{\left(4\pi s\right)^{2}}\int\sqrt{-g}\, d^{4}x+\mathcal{O}\left(s^{-1}\right).\label{eq:TrExp}
\end{equation}
Then
\begin{equation}
S_{\textrm{eff}}^{\Lambda}=\mp\frac{1}{2}\frac{\Lambda^{4}}{2}\frac{1}{\left(4\pi\right)^{2}}\int\sqrt{-g}\, d^{4}x+\mathcal{O}\left(\Lambda^{2}\right).\label{eq:SeffL}
\end{equation}

\par\end{flushleft}

Expressing (temporary, i.e.\ at the moment $t=0$) gravitational
field as a sum of the classical (Minkowski) background $\eta_{\mu\nu}$
and the ``quantum'' field $h_{\mu\nu}$,\ i.e. $g_{\mu\nu}$=$\eta_{\mu\nu}+\kappa h_{\mu\nu},$
we can perform the following expansion
\begin{equation}
\sqrt{-g}=1+\frac{1}{2}\kappa h_{\mu}^{\mu}+\frac{1}{8}\kappa^{2}h_{\mu}^{\mu}h_{\nu}^{\nu}-\frac{1}{4}\kappa^{2}h_{\mu\nu}h^{\mu\nu}+\ldots.\label{eq:detgexpansion}
\end{equation}
For the flat Friedmann--Lema�tre--Robertson--Walker (FLRW) cosmological
model the only non-zero components of the quantum field are given
by $\kappa h_{ii}=1-a^{2}\left(t\right)\equiv\kappa b\left(t\right)$,
where $a\left(t\right)$ is the scale parameter of the FLRW metric.
The first term in $\left(\ref{eq:detgexpansion}\right)$, i.e.\ $\textrm{constant}=1$,
does not give any contribution to equations of motion, and it can
be ignored (it does not couple to gravitational field). Alternatively,
we can remove it because of a standard normalization of the effective
action following from the subtraction corresponding to $\mathcal{D}_{0}$
($\mathcal{D}_{0}\equiv\mathcal{D}\left(g_{\mu\nu}=\eta_{\mu\nu}\right)$)
and/or of a normalization of functional measure. The second term,
linear in $h_{\mu\nu}$, is a tadpole contribution. Its explicit form
in terms of $b$ is
\begin{equation}
\frac{1}{2}\kappa h_{\mu}^{\mu}=-\frac{3}{2}\kappa b=\frac{3}{2}\left(a^{2}-1\right)=\frac{3}{2}\left(2\dot{a}t+\ddot{a}t^{2}+\dot{a}^{2}t^{2}+\ldots\right),\label{eq:bexp}
\end{equation}
where the expansion around $t=0$ has been introduced, i.e.\ $a\left(t\right)=a\left(0\right)+\dot{a}t+\frac{1}{2}\ddot{a}t^{2}+\ldots$,
with $a\left(0\right)=1$. The term linear in $t$ can be gauged away
in $\left(\ref{eq:bexp}\right)$ ($g_{\mu\nu}\rightarrow g_{\mu\nu}+\partial_{\mu}\xi_{\nu}+\partial_{\nu}\xi_{\mu}$)
with the gauge function parameter
\begin{equation}
\xi_{\mu}=\left(\frac{1}{2}H\mathbf{x}^{2},-Ht\mathbf{x}\right).\label{eq:gaugepar}
\end{equation}
Moreover, since $h_{\mu\nu}\propto b=\mathcal{O}\left(t^{2}\right)$,
further terms in $\left(\ref{eq:detgexpansion}\right)$ are of the
order $\mathcal{O}\left(t^{4}\right)$ and will be ignored as they
would give a negligible contribution to the averaging procedure $\left(\ref{eq:meanT}\right)$.
In terms of the Hubble parameter we have

\noindent \begin{flushleft}
\begin{equation}
\sqrt{-g}-1\sim\frac{3}{2}\left(\ddot{a}t^{2}+\dot{a}^{2}t^{2}\right)=\frac{3}{2}\left(\dot{H}+2H^{2}\right)t^{2}.\label{eq:expg}
\end{equation}
Finally,
\begin{equation}
S_{\textrm{eff}}^{\Lambda}\approx\mp\frac{3}{128\pi^{2}}\Lambda^{4}V_{3}\int t^{2}dt\left(\dot{H}+2H^{2}\right).\label{eq:Seff-final}
\end{equation}

\par\end{flushleft}

Here $V_{3}$ is a three-dimensional volume, and integration with
respect $t$ requires a special treatment. Since we are interested
in the energy density, we should divide (\ref{eq:Seff-final}) by
$-t\times V^{3}$, where the minus sign follows from the relation
between the energy density and the Lagrangian density. Namely, $\varrho\thickapprox-\mathcal{L}$,
because for slowly changing fields (in the scale of the Planck time
$t_{\textrm{Planck}}$), which certainly takes place in cosmology,
kinetic part (velocities) can be skipped. Dividing by $V^{3}$ is
obvious, but dividing the $t$-integral by $t$ is an averaging procedure
with respect to $t$. To be within perturbative (in $t$) regime we
should consider the finest average, which is attained for the shortest
physically possible time, i.e.\ $t_{\textrm{Planck}}$. The average
reads 
\begin{equation}
\frac{1}{t}\intop t^{2}dt\rightarrow\left.\left(\frac{1}{t}\int t^{2}dt\right)\right|_{0}^{t_{\textrm{Planck}}}=\frac{1}{3}t_{\textrm{Planck}}^{2}\approx\frac{1}{3}G.\label{eq:meanT}
\end{equation}
We have omitted in the averaging procedure $\left(\ref{eq:meanT}\right)$
the expression in the parenthesis in Eq.~$\left(\ref{eq:Seff-final}\right)$
because $H$ is practically constant within the Planck time. Then
\begin{equation}
\varrho_{\textrm{d}}\approx\pm\frac{G}{128\pi^{2}}\Lambda^{4}\left(\dot{H}+2H^{2}\right).\label{eq:rod}
\end{equation}

Assuming the Planck UV cutoff ($\Lambda\thickapprox\nicefrac{1}{\sqrt{G}}$,),
we obtain the DE density
\begin{equation}
\varrho_{\textrm{d}}\approx\pm\frac{1}{16\pi\kappa}\left(\dot{H}+2H^{2}\right)\label{eq:rodfinal}
\end{equation}
per mode. It appears that $\left(\ref{eq:rodfinal}\right)$ exactly
expresses the form of the Ricci DE density (\ref{eq:defroD}). Thus,
we have presented a {}``microscopic'' derivation of the Ricci DE
model.

The consequences of such a form of the DE density are well-known.
The Friedmann equations with the Ricci DE density (\ref{eq:defroD})
and with the matter density $\varrho_{\textrm{m}}$ are easily solvable.
In fact, the only equation is
\begin{equation}
H^{2}=\frac{\kappa}{3}\varrho\equiv\frac{\kappa}{3}\left(\varrho_{\textrm{d}}+\varrho_{\textrm{m}}\right)\equiv\alpha\left(\dot{H}+2H^{2}\right)+\frac{\kappa}{3}\varrho_{\textrm{m}},\label{eq:H2eqro}
\end{equation}
where (\ref{eq:defroD}) has been inserted. A significant technical
simplification of (\ref{eq:H2eqro}) is achieved with help of a new
variable $x\equiv\log a$. Namely
\begin{equation}
H^{2}=\alpha\left(\frac{1}{2}\frac{dH^{2}}{dx}+2H^{2}\right)+\frac{\kappa}{3}\varrho_{\textrm{m}0}e^{-3x},\label{eq:H2final}
\end{equation}
where the standard form of $\varrho_{\textrm{m }}$ has been explicitly
introduced. The general solution of the linear first order Eq.~\ref{eq:H2final}
is given by
\begin{equation}
H^{2}\left(x\right)=\frac{2\kappa}{3\left(2-\alpha\right)}\varrho_{\textrm{m0}}e^{-3x}+Ce^{-\left(4-\frac{2}{\alpha}\right)x},\label{eq:H2sol}
\end{equation}
with the constant $C$ determined by initial conditions (for $x=0$).

It would be very interesting to confront the theoretically estimated
coefficient in front of (\ref{eq:rodfinal}), corresponding to $\alpha$
in (\ref{eq:defroD}), with observational parameter $\alpha_{\textrm{Obs}}$
determined by data fitting in the framework of the cosmological evolution
dictated by $\left(\ref{eq:H2sol}\right)$. For example, according
to \citet{Wang:2011km} the best fits for $\alpha_{\textrm{Obs}}$
are $\alpha_{\textrm{W}}\approx0.23$ and $\alpha_{\textrm{WBS}}\approx0.35$,
where the first case corresponds to WMAP data alone, and the second
one to WMAP+BAO+SNIa data, respectively. According to (\ref{eq:rodfinal})
our $\alpha$ per mode (following from quantum field theory) is $\alpha_{\textrm{QFT}}\approx1/3\cdot16\pi\approx0.0066$.
Then, the fraction $N\equiv\alpha_{\textrm{Obs}}/\alpha_{\textrm{QFT}}\approx35\div53\sim50$
is the effective expected number of fundamental particle modes necessary
to fit the theoretical form of the Ricci DE to observational data.
We have rounded the fraction $N$ to a single number because, due
to the two approximations (approximate status of the average $\left(\ref{eq:meanT}\right)$,
and the approximate value of the UV cutoff), we know the value of
$\alpha_{\textrm{QFT }}$ imprecisely.

In conclusion, we would like to stress that it is possible to {}``microscopically''
derive the Ricci DE model in the framework of quantum field theory
of {}``matter'' fields in a classical gravitational background.
Moreover, we have found that data fitting reasonably requires the
effective number of fundamental particle modes of the order of 50.

\end{document}